\documentstyle[12pt]{article}

\begin{document}

\title{{\bf Electromagnetic Transition in Waveguide with Application to
Lasers}}
\author{Dai, Wu-Sheng$^{1}$ and Xie, Mi$^{2}$ \\
{\footnotesize $^{1}$Department of Applied Physics, Tianjin University,
Tianjin 300072, P. R. China}\\
{\footnotesize $^{2}$Department of Physics, Tianjin Normal University,
Tianjin 300074, P. R. China}}
\date{}
\maketitle

\begin{abstract}
The electromagnetic transition of two-level atomic systems in a waveguide is
calculated. Compared with the result in free space, the spontaneous emission
rate decrease because the phase space is smaller, and meanwhile, some
resonance appears in some cases. Moreover, the influence of non-uniform
electromagnetic field in a waveguide on absorption and stimulated emission 
is considered. Applying the results to lasers, a method to enhance the laser
power is proposed.
\end{abstract}

PACS codes: 42.50.ct, 42.55.Ah

{\noindent {\bf 1 Introduction}}

The behavior of atoms in confined space had been studied for many years
since the possibility of modifying spontaneous emission rates was first
mentioned by Purcell \cite{purcell}. Many studies had been done in
theoretical and experimental aspects \cite{barton,meystre,berman}. The
transition rate of atoms may increase \cite{purcell} or decrease \cite
{kleppner} in a cave. In this paper, we will discuss spontaneous emission,
absorption and stimulated emission of atoms in a waveguide and apply the
results to the problem of how to enhance the output power of lasers.

In confined space, the values of momenta are discrete at certain directions
and a nonzero lower limit exists due to the boundaries. Compared with free
space, the phase space in confined space is smaller. Some processes, {\it %
e.g.} atomic decay, will be influenced. In the case of spontaneous emission,
the transition rate of excited atoms is determined by two factors: The
transition matrix element between two states and the phase space volume of
final state. In confined space, the transition matrix element is the same as
in free space under the first order approximation. The phase space volume is
smaller because some components of photon momenta in final state are
constrained to some discrete values. It leads to the decrease of the total
spontaneous emission rate of excited atoms. On the other hand, when the
frequency of the photon in final state equals to one of the eigenfrequencies
of the cave, resonance will occur and the transition rate will remarkably
increase. In this paper, we will give a direct calculation about the
spontaneous emission rate of excited atoms in a matrix waveguide. Also, the
ratio of the spontaneous emission rate in the waveguide to the case in free
space is given. The ratio is determined only by the shape of the boundary,
but it is independent of the form of the transition matrix. This means that
the ratio is not related to special transition processes. Besides, since the
electromagnetic field has a defined non-uniform distribution in a waveguide,
the rates of absorption and stimulated emission will depend on the position
of the atoms. The corresponding transition rate at each position in the
waveguide and the mean transition rate are also given.

The power of lasers is related to the spontaneous-emission lifetime of the
excited states. The lifetime of atoms can be influenced in confined space.
Therefore, we can change the laser power by putting the atoms in a cavity.
In the last part of this paper, we will discuss the change of the output
power of lasers under the influence of the scale of waveguides. 

\vskip 5mm

{\noindent {\bf 2 Spontaneous emission in a waveguide}}

\vskip 5mm 

For one-photon decay of an initial state $|a\rangle $ into a final
state $|b\rangle $ and a photon with frequency $\omega _{n}$ and
polarization $\sigma $, the $S$-matrix element to first order in
perturbation theory is (Here we use Heaviside's units and take $c=\hbar =1$) 
\cite{gross}: 
\begin{equation}
S_{ba}=-i2\pi \delta (E_{b}+\omega _{n}-E_{a})\frac{1}{\sqrt{L_{x}L_{y}L_{z}}%
}f_{ba},  \label{e2}
\end{equation}
where $E_{a}$ and $E_{b}$ are the energies of the two states of the atomic
system. $f_{ba}$ is the transition amplitude. Under the dipole
approximation, the amplitude takes the form: 
\begin{equation}
f_{ba}=-ie\sqrt{\frac{\omega _{n}}{2}}{\bf \epsilon }_{n}^{\sigma \ast
}\cdot {\bf r}_{ba}.  \label{e3}
\end{equation}
The vectors ${\bf \epsilon }$ are polarization vectors and ${\bf r}_{ba}$ is
the matrix element 
\begin{equation}
{\bf r}_{ba}=\langle b|{\bf r}|a\rangle .  \label{e4}
\end{equation}

Consider a two-level system placed in a waveguide with sides $L_{x}$ and $%
L_{y}$ which are comparable with the wavelength of the photon. In the
waveguide, the components of the momentum at $x$ and $y$ directions of the
photon are constrained to some discrete values: 
\begin{equation}
\begin{array}{c}
\displaystyle k_{x}=\frac{n_{x}\pi }{L_{x}},~~~~~~k_{y}=\frac{n_{y}\pi }{%
L_{y}},~~~~~~~~~(n_{x},n_{y}=0,1,2,\cdots ).
\end{array}
\label{e8}
\end{equation}
Summing over all final photon states, the total spontaneous emission rate is 
\begin{equation}
W_{sp}=\sum\limits_{n_{x}n_{y}n_{z}}\frac{1}{L_{x}L_{y}L_{z}}2\pi \delta
(E_{b}+\omega _{n}-E_{a})|f_{ba}|^{2}.  \label{e9}
\end{equation}
Since the photon is free at $z$ direction, the summation of $n_{z}$ can be
replaced by integration: 
\begin{equation}
\begin{array}{lll}
W_{sp} & = & \displaystyle\sum\limits_{n_{x}n_{y}}\frac{1}{L_{x}L_{y}}\int 
\frac{dk_{z}}{2\pi }2\pi \delta (E_{b}+\omega _{n}-E_{a})|f_{ba}|^{2} \\[2mm]
& = & \displaystyle\sum\limits_{n_{x}n_{y}}\frac{1}{L_{x}L_{y}}\frac{\omega 
}{k_{0}}|f_{ba}|^{2} \\[2mm]
& = & \displaystyle\sum\limits_{n_{x}n_{y}}\frac{1}{L_{x}L_{y}}\frac{%
e^{2}\omega ^{2}}{2k_{0}}|{\bf r}_{ba}|^{2}\sin ^{2}{\theta },
\end{array}
\label{e10}
\end{equation}
where $k_{0}$ is the $z$-component of wave vector ${\bf k}$ 
\begin{equation}
k_{0}=\sqrt{\omega ^{2}-k_{x}^{2}-k_{y}^{2}}  \label{e10a}
\end{equation}
and $\theta $ is the angle between the dipole moment\ and wave vector ${\bf k%
}$.

In free space, the total spontaneous emission rate is known as \cite{gross}

\begin{equation}
W_{sp}^{free}=\frac{e^{2}\omega ^{3}}{3\pi }|{\bf r}_{ba}|^{2}.  \label{e10b}
\end{equation}
Comparing with eq.(\ref{e10}), we obtain the ratio of the rates in confined
space and free space: 
\begin{equation}
\eta =\frac{W_{sp}}{W_{sp}^{free}}=\sum\limits_{n_{x}n_{y}}\frac{1}{%
L_{x}L_{y}}\frac{3\pi }{2\omega k_{0}}\sin ^{2}\theta .  \label{e11}
\end{equation}
Because the space in a waveguide is not homogeneous, a new factor $\sin
^{2}\theta $ is appeared. This result shows that the spontaneous emission
rate is different if the polarization of atoms in the waveguide is
different. However, atoms usually are non-polarized, so a mean ratio is
needed.

Use ($\alpha $, $\beta $) to describe the direction of ${\bf r}_{ba}$ in the
waveguide. Since 
\begin{equation}
\begin{array}{lll}
\cos \theta & = & \displaystyle\frac{{\bf k}\cdot {\bf e}_{r}}{|{\bf k}|} \\%
[2mm] 
& = & \displaystyle\frac{k_{x}\sin \alpha \cos \beta +k_{y}\sin \alpha \sin
\beta +k_{z}\cos \alpha }{|{\bf k}|},
\end{array}
\label{e12}
\end{equation}
the mean ratio can be obtained as 
\begin{equation}
\begin{array}{lll}
\bar{\eta} & = & \displaystyle\frac{1}{4\pi }\int \eta ~d\Omega \\[2mm] 
& = & \displaystyle\sum\limits_{n_{x}n_{y}}\frac{1}{L_{x}L_{y}}\frac{\pi }{%
\omega k_{0}}.
\end{array}
\label{e13}
\end{equation}
Obviously, this result is only related with three length scales: the
wavelength of the photon $\lambda $, the widths of the waveguide $L_{x}$ and 
$L_{y}$. Introducing two parameters: 
\begin{equation}
\gamma _{x}=\frac{L_{x}}{\lambda /2},~~~~~~~~~~~~~\gamma _{y}=\frac{L_{y}}{%
\lambda /2},  \label{e14}
\end{equation}
we can get the following expression: 
\begin{equation}
\bar{\eta}=\sum\limits_{n_{x}n_{y}}\frac{1}{\pi }\frac{1}{\sqrt{\gamma
_{x}^{2}\gamma _{y}^{2}-\gamma _{y}^{2}n_{x}^{2}-\gamma _{x}^{2}n_{y}^{2}}}.
\label{e15}
\end{equation}

This result shows that the spontaneous emission rate is usually smaller in a
waveguide than in free space because the phase space of the final state is
smaller and the possible states of photon are fewer. However, the
spontaneous emission rate will diverge when $\gamma _{x}$ and $\gamma _{y}$
take such values that the denominator of (\ref{e15}) is zero. From (\ref
{e10a}) we can see that this is just the case that the $z$-component of
photon $k_{z}$ or $k_{0}$ vanishes. In other words, when $k_{0}=0$, some
resonance occurs and the spontaneous emission rate of excited atoms
increases rapidly (see fig. 1).

Besides, when both $\gamma _{x}$ and $\gamma _{y}$ are smaller than 1 (see
fig. 2), the frequency of the photon is lower than the lowest permitted
frequency in such a waveguide. It shows that the phase space of the final
state is compressed to zero so the photon can not be emitted and the
transition can not occur. At that time, an excited atom will never return to
its ground state, {\it i.e.}, the transition is forbidden.

\vskip 5mm

{\noindent {\bf 3 Absorption and stimulated emission in a waveguide}}

\vskip 5mm

The rates of absorption and stimulated emission in confined space are
different from those in free space because the mode structures of
electromagnetic fields are different in the two cases. The processes of
absorption and stimulated emission are transitions of atomic systems in
external electromagnetic fields. The electromagnetic field in a waveguide
has a defined distribution which is not uniform, so the transition
probabilities will depend on the position of the atoms.

In a matrix waveguide, the expressions for the spatial components of the
field are 
\begin{equation}
\left\{ 
\begin{array}{l}
E_{x}(x,y,z)=E_{1}\cos k_{x}x\sin k_{y}ye^{ik_{z}z}, \\ 
E_{y}(x,y,z)=E_{2}\sin k_{x}x\cos k_{y}ye^{ik_{z}z}, \\ 
E_{z}(x,y,z)=E_{3}\sin k_{x}x\sin k_{y}ye^{ik_{z}z},
\end{array}
\right.
\end{equation}
where $k_{x},k_{y},k_{z}(=k_{0})$ have been given in (\ref{e8}) and (\ref
{e10a}), and $E_{1},E_{2},E_{3}$ satisfy $%
k_{x}E_{1}+k_{y}E_{2}-ik_{z}E_{3}=0 $.

We will first consider the processes of absorption in the waveguide. In this
case, a monochromatic electromagnetic wave at frequency $\omega $ is made to
interact with an atom which is in state $|b\rangle $. Since the magnitudes
of the atomic electron radii are much smaller than the wavelength of the
electromagnetic wave, the dipole approximation is extremely good. On this
assumption, we regard ${\bf E}$ as an uniform field in a small region and
use its value at $z=0$. The corresponding electric field is a sinusoidal
function of time ${\bf E}(x,y,t)={\bf E}_{0}(x,y)\sin \omega t$, where ${\bf %
E}_{0}(x,y)=E_{x}(x,y,0){\bf i}+E_{y}(x,y,0){\bf j}+E_{z}(x,y,0){\bf k}$.
Similar to the case of absorption occurring in free space \cite{svelto}, the
rate of absorption can be obtained directly: 
\begin{equation}
W_{ab}(x,y)=\frac{\pi }{3n^{2}}|{\bf r}_{ab}|^{2}\rho (x,y)\delta (\Delta
\omega ).
\end{equation}
Here $n$ is the refractive index of the atomic system; $\Delta \omega $ is
the magnitude of the spread in $\omega $; ${\bf r}_{ab}$ is the matrix
element corresponding to the transition from the initial state $|b\rangle $
to the final state $|a\rangle $; $\rho $ is the energy density of the
incident electromagnetic wave in the waveguide:

\begin{eqnarray}
\rho (x,y) &=&\frac{1}{2}n^{2}E_{0}(x,y)^{2} \\
&=&\frac{1}{2}n^{2}(E_{1}^{2}\cos ^{2}k_{x}x\sin ^{2}k_{y}y+E_{2}^{2}\sin
^{2}k_{x}x\cos ^{2}k_{y}y+E_{3}^{2}\sin ^{2}k_{x}x\sin ^{2}k_{y}y). 
\nonumber
\end{eqnarray}
Obviously, the transition probability depends on the position of the atom,
since the electromagnetic field in waveguide has a non-uniform distribution.
The mean rate of absorption in the waveguide is

\begin{equation}
\overline{W}_{ba}=\frac{\pi }{3n^{2}}|{\bf r}_{ba}|^{2}\bar{\rho}\delta
(\Delta \omega ),
\end{equation}
with the mean energy density

\begin{equation}
\bar{\rho}=\frac{1}{V}\int_{V}\rho (x,y)dV=\frac{n^{2}}{8}%
(E_{1}^{2}+E_{2}^{2}+E_{3}^{2}).
\end{equation}

The rate of stimulated emission $W_{ba}$ is equal to the rate of absorption $%
W_{ab}$, since the transition matrix element $|{\bf r}_{ba}|=|{\bf r}_{ab}|$
and the photons of stimulated emission must always match the mode of the
stimulating photons.

\vskip 5mm

{\noindent {\bf 4 Application to lasers}}

\vskip 5mm

One of the central problems in the area of laser is how to enhance the laser
power. There are many factors which can influence the laser power. One of
them is the spontaneous-emitted lifetime of excited states. According to the
results above, the spontaneous emission rate of excited atoms in confined
space, {\it e.g.} waveguide, is different to the case of free space. This
means that we can change the lifetime of excited atoms by adjusting the
shape and scale of the cavity (In this paper, the cavity is a waveguide).
One factor which directly influences the laser power is the lifetime of the
upper laser level. If we put the laser medium in a waveguide, we can enhance
or reduce the laser power by changing the lifetime of the upper laser level.

Take a three-level laser proceeds as an example, the output power of
continuous wave laser under the rate-equation approximation is \cite{svelto} 
\begin{equation}
P=\displaystyle A\frac{\chi -1}{\tau }.  \label{e16}
\end{equation}
Here $A$ is a parameter determined by the character of the specific laser
medium and laser facility. The quantity $\tau $ is the lifetime of the upper
laser level, and it is, in general, given by

\begin{equation}
\frac{1}{\tau }=\frac{1}{\tau _{sp}}+\frac{1}{\tau _{nr}},  \label{e17}
\end{equation}
where $\tau _{sp}$ is the spontaneous-emission lifetime and $\tau _{nr}$ the
nonradiative lifetime of upper laser level. Let $W_{p}$ be the pumping rate
and $W_{cp}$ the critical pumping rate. The $\chi $ is the amount by which
threshold is exceeded:

\begin{equation}
\chi =\frac{W_{p}}{W_{cp}}.  \label{e18}
\end{equation}
In practice, the following approximation can be taken:

\begin{equation}
W_{p}\simeq \frac{1}{\tau }.  \label{e19}
\end{equation}
Then the output power is given as

\begin{equation}
P=\displaystyle A(W_{p}-\frac{1}{\tau _{nr}}-\frac{1}{\tau _{sp}}).
\label{e20}
\end{equation}

If the process occurs in a waveguide, $\tau _{sp}$ will be replaced by $\tau
_{sp}^{W}$ . The $\tau _{sp}^{W}$ is the spontaneous-emission lifetime in
the waveguide. From eq.(\ref{e11}), we can obtain

\begin{equation}
\eta =\frac{W_{sp}}{W_{sp}^{free}}=\frac{1/\tau _{sp}^{W}}{1/\tau _{sp}^{{}}}%
=\frac{\tau _{sp}^{{}}}{\tau _{sp}^{W}}.  \label{e21}
\end{equation}
Substitute this relation into eq.(\ref{e20}), the laser power in a waveguide 
$P^{W}$ is therefore 
\begin{equation}
P^{W}=\displaystyle A(W_{p}-\frac{1}{\tau _{nr}}-\frac{\eta }{\tau _{sp}}).
\label{e22}
\end{equation}
When $\eta <1$, the output power enhances, {\it i.e.} $P^{W}>P$. On the
contrary, when $\eta >1$, the output power reduces, {\it i.e.} $P^{W}<P$.
The analysis carried out here implies that the output power can be adjusted
by changing the scale of the waveguide.

Besides, in waveguides, by adjusting the spontaneous-emission lifetime of
the upper laser level $\tau _{sp}^{W}$, we can obtain a high power laser
pulse. If $\tau _{sp}^{W}$ is long enough, after sufficient pumping, a large
population inversion will be obtained. For instance, the spontaneous
emission of the upper laser level can be forbidden to a certain extent by
modifying the scale of the waveguide. When the population inversion is large
enough, change the waveguide to a proper scale to induce a spontaneous
transition of the upper laser level. In this way, a laser pulse is obtained.

In conclusion, we have shown that transition rates of atoms may change in
confined space. In some cases, the lifetime can be changed by adjusting the
experimental arrangements. Based on this idea, a method to enhance the laser
power is proposed.

\newpage \vskip 3cm \noindent {\bf Figure captions:}\vskip 1cm

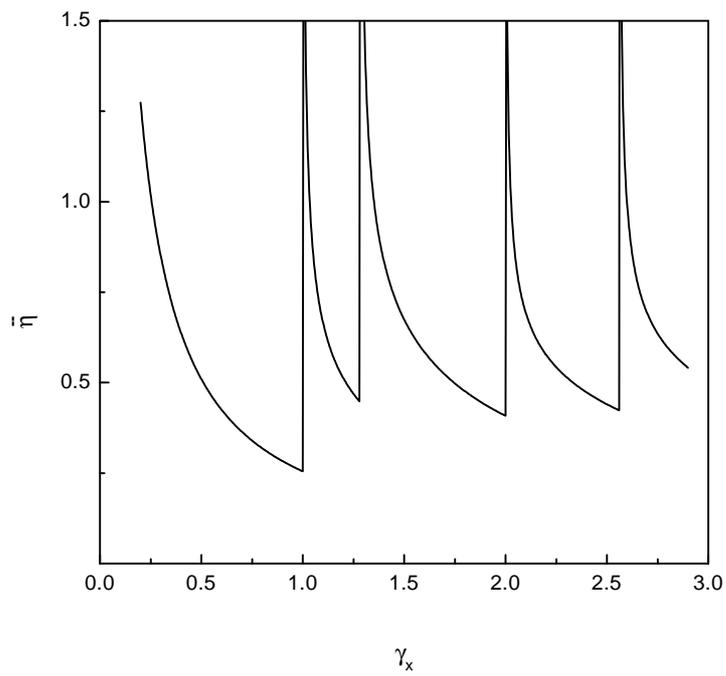
\noindent Fig. 1. In a waveguide, the decay rate is usually depressed since
the phase space is smaller than in free space, but remarkably increases when
resonance occurs ($\gamma_y=1.6$). \vskip 1cm

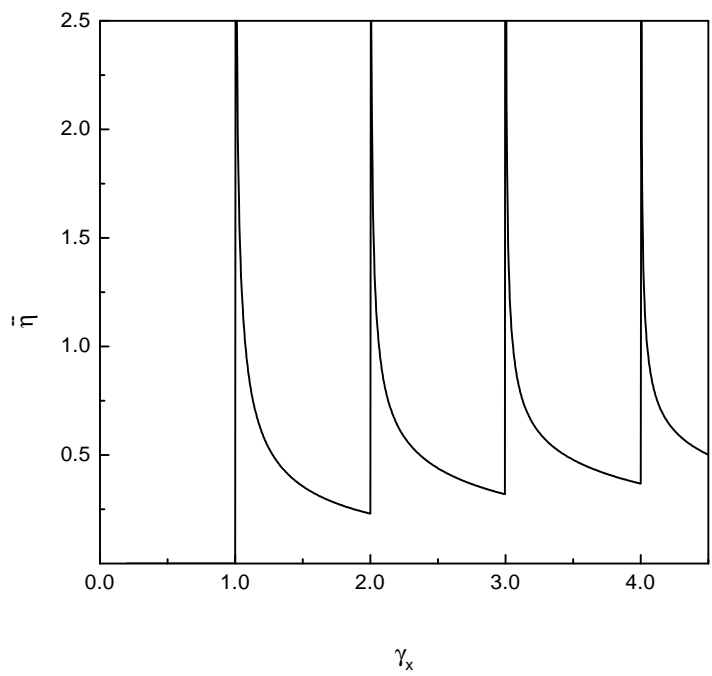
\noindent Fig. 2. When $\gamma_x,\gamma_y<1$, the decay is forbidden (Here $%
\gamma_y=0.8$).

\end{document}